# CUBES, THE CASSEGRAIN U-BAND EFFICIENT SPECTROGRAPH: TOWARDS FINAL DESIGN REVIEW.

Matteo Genoni[1,*], Hans Dekker[1], Stefano Covino[1], Roberto Cirami[2], Marcello A. Scalera[1], Lawrence Bissel[3], Walter Seifert[4], Ariadna Calcines[5], Gerardo Avila[6], Marco Landoni[1], Julian Störmer[4], Christopher Ritz[4], David Lunney[3], Chris Miller[3], Stephen Watson[3], Chris Waring[3], Bruno V. Castilho[7], Marcio V. De Arruda[7], Orlando Verducci[7], Igor Coretti[2], Luca Oggioni[1], Giorgio Pariani[1], Edoardo A. M. Redaelli[1], Matteo D' Ambrogio[1], Giorgio Calderone[2], Matteo Porru[2], Ingo Stilz[4], Rodolfo Smiljanic[9], Guido Cupani[2], Mariagrazia Franchini[2], Andrea Scaudo[1], Vincent Geers[3], Vincenzo De Caprio[10], Domenico D' Auria[10], Mina Sibalic[10], Cyrielle Opitom[11], Gabriele Cescutti[12], Valentina D' Odorico[2], Ruben Sanchez-Janssen[3], Andreas Quirrenbach[4], Beatriz Barbuy[8], Stefano Cristiani[2], Paolo Di Marcantonio[2].

[1]INAF-Osservatorio Astronomico di Brera, via E. Bianchi 46, 23807 Merate, Italy;
[2]INAF-Osservatorio Astronomico di Trieste, via G. B. Tiepolo 11, 34131 Trieste, Italy;
[3]STFC - United Kingdom Astronomy Technology Centre (UK ATC), Edinburgh, UK;
[4]Landessternwarte - Zentrum für Astronomie der Universitat Heidelberg, Heidelberg, Germany;
[5]Durham University - Centre for Advanced Instrumentation - Department of Physics, UK;
[6]Optical Lab Baader Planetarium GmbH, Zur Sternwarte 4D-82291 Mammendorf Germany;
[7]LNA/MCTI, Rua Estados Unidos, 154 - 37504-364, Itajubá, Brazil;
[8]Universidade de São Paulo - IAG, Rua do Matão 1226, São Paulo, Brazil;
[9]Nicolaus Copernicus Astronomical Center - Polish Academy of Sciences, Warsaw, Poland;
[10]INAF-Osservatorio Astronomico di Capodimonte, via Moiariello 16, 80131 Napoli, Italy;
[11]Institute for Astronomy, University of Edinburgh, Royal Observatory, Edinburgh, UK;
[12]Dipartimento di Fisica-Sezione di Astronomia Università di Trieste, Trieste, Italy;

## ABSTRACT

In the era of Extremely Large Telescopes, the current generation of 8-10m facilities are likely to remain competitive at ground-UV wavelengths for the foreseeable future. The Cassegrain U-Band Efficient Spectrograph (CUBES) has been designed to provide high instrumental efficiency ( $> 37\%$) observations in the near UV (305-400 nm requirement, 300-420 nm goal) at a spectral resolving power of R $> 20,000$ (with a lower-resolution, sky-limited mode of R $\sim 7,000$). With the design focusing on maximizing the instrument throughput (ensuring a Signal to Noise Ratio – SNR– $\sim 20$ per spectral resolution element at 313 nm for U $\sim 17.5$ mag objects in 1h of observations), it will offer new possibilities in many fields of astrophysics: i) access to key lines of stellar spectra (e.g. lighter elements, in particular Beryllium), extragalactic studies (e.g. circumgalactic medium of distant galaxies, cosmic UV background) and follow-up of explosive transients. We present the CUBES instrument design, currently in Phase-C and approaching the final design review, summarizing the hardware architecture and interfaces between the different subsystems as well as the relevant technical requirements. We describe the optical, mechanical, electrical design of the different subsystems (from the telescope adapter and support structure, through the main opto-mechanical path, including calibration unit, detector devices and cryostat control, main control electronics), detailing peculiar instrument functions like the Active Flexure Compensation (AFC). Furthermore, we outline the AIT/V concept and the main instrument operations giving an overview of its software ecosystem. Installation at the VLT is planned for 2028/2029 and first science operations in late 2029.

*Contact Matteo Genoni email: matteo.genoni@inaf.it

# 1. INTRODUCTION

The four 8.2m telescopes of the Very Large Telescope (VLT) at the European Southern Observatory (ESO) are the world's most scientifically productive ground-based observatory in the visible and infrared. Looking to the future of the VLT there is a long-standing aspiration for an optimized ultraviolet (UV) spectrograph[1].

The European Extremely Large Telescope (ELT), under construction in northern Chile by ESO, with a primary aperture of 39m will be unprecedented in its light-gathering power, coupled with exquisite angular resolution via correction for atmospheric turbulence by adaptive optics (AO). The choice of protected silver (Ag+Al) for the ELT mirror coatings (excl. M4) ensures a durable, proven surface with excellent performance across a wide wavelength range, but the performance drops significantly in the blue-visible part of the spectrum compared to bare aluminium. ESO is actively researching alternative coatings, but in the short-medium term we can assume that the performance of the ELT in the blue-visible (<450 nm) will be limited in this regard.

Therefore, a blue-optimized instrument on the VLT will be competitive and complementary to the ELT at wavelengths shorter than 400 nm. In January 2020 ESO issued a Call for Proposal for a Phase A study of a UV Spectrograph to be installed at a Cassegrain focus of the VLT, with the goal of high-efficiency( > 37/40%) and intermediate resolving power (~ 20k) in the ground-UV domain (305-400 nm requirement, 300-420 nm goal).

In May 2020 the CUBES (Cassegrain U-Band Efficient Spectrograph) Consortium, led by INAF, was selected to carry out the study. The CUBES project completed a Phase A conceptual design in June 2021. After the endorsement by the ESO Council at the end of 2021, Phase B started in February 2022 with the signature of the Construction Agreement between ESO and the leading institute of the CUBES Consortium, opening the detailed design and construction phase. Here we report the present status of the project, which will provide a world-leading UV capability for ESO from 2028 well into the ELT era.

# 2. CUBES KEY SCIENCE CASES AND TOP LEVEL REQUIREMENTS

The CUBES science case spans a broad range of contemporary astrophysics. Here we highlight key cases across Solar System, Galactic, and extragalactic science that are driving the design.

## 2.1 Stellar nucleosynthesis

More than a quarter of the chemical elements spectral features are only observable in the near UV. The low efficiency of instruments in this domain, however, severely restricted previous studies. Advancements in the field require high-resolution, near-UV spectroscopy of a large number and diversity of stars. CUBES allows us to deal with a few fundamental problems:

- Metal-poor stars and light elements. A key case is to probe the early chemical evolution of the Galaxy, via chemical abundance patterns in the oldest, low-mass stars that formed from material enriched by the first supernovae. The so-called Carbon-enhanced metal-poor stars are the perfect probes to investigate nucleosynthesis by the first stars. CUBES will enable quantitative spectroscopy for large samples of metal-poor stars, providing direct estimates for a broad range of heavy elements, as well as valuable constraints on CNO elements.

- Heavy-element nucleosynthesis. Stellar abundances from CUBES will provide critical tests of the various production channels of heavy elements for both r- and s-process elements. Determining the abundances of neutron-capture elements in metal-poor stars is fundamental to understanding the physics of these processes and the chemical evolution of the Galaxy as well as the origin of the Galactic halo.

- Beryllium is one of the lightest and simplest elements. Nevertheless, questions remain about its production in the early Universe. Recent results are consistent with no primordial production, but larger samples are required to investigate this further. CUBES will provide large homogeneous samples of Be abundances in stars belonging to different populations providing new insights into its production and tracing the star-formation history of the Galaxy.

## 2.2 Extragalactic studies and Cosmology

- Galaxies are likely able to produce most of the UV emissivity needed for cosmic reionisation at high redshift but quasars also possibly contribute. Estimates of the escape fraction $f_{esc}$ of hydrogen-ionising photons able to escape a galaxy are close to 100\% for quasars. However, the volume density of low- and intermediate-luminosity quasars at z > 4 is still uncertain, so it is unclear if they are the dominant source of ionisation. In contrast, star-forming galaxies are more numerous, but estimates of $f_{esc}$ from observations of the Lyman continuum have uncertainties of tens of percent and are limited to a handful of systems at z = 2.5 to 4. To be detectable from Earth escaping photons must survive absorption along the line of sight by the intergalactic medium, which become stronger with redshift and is significantly variable between sightlines. Given these competing factors, the ideal redshift range for ground-based observations of the Lyman continuum of a galaxy is z = 2.4 to 3.5, i.e. from about 410nm down to the atmospheric cut-off. For this reason, CUBES could be an asset for this science case thanks to its high throughput. Furthermore, since the galaxies to be observed are extremely faint this science case is also one of the main drivers of the low-resolution mode.

- Within the Standard Model of particle physics and cosmology there is still no accepted model for dark energy and dark matter, or why the Universe contains baryons instead of antibaryons, or even why the Universe contains baryons at all. We are also missing crucial properties of neutrinos (e.g. their hierarchy, why they change flavour, and the number that existed during the earliest phases of the Universe). Some of these questions can be investigated by measuring the nuclides produced a few minutes after the Big Bang. The primordial deuterium (D/H) abundance is currently our most reliable probe of Big Bang Nucleosynthesis. CUBES will provide a large, reliable sample of D/H estimates from quasar absorption spectra.

- Remarkable progresses about the missing baryon problem have been recently made possible at low redshifts by studying the dispersion measure in Fast Radio Bursts (FRBs) and at z > 1.5 by observations and simulations of the Lyman forest. Still, we have insufficient knowledge about how baryonic matter is distributed among the different gaseous components and we would need to better constrain the mechanisms (stellar and AGN feedback, accretion, etc.) that determine the observed distribution. A UV efficient spectrograph with relatively high resolution offers the possibility to dig into the complex nature of the inter- and circum-galactic gas at z ~ 1.5 to 3, via observations of quasar absorption lines.

## 2.3 Follow-up of explosive transients

Time-domain astronomy is one of the most active branches of modern astrophysics. In a few years, new observational facilities, specifically designed with the goal of securing high-cadence observations of large fractions of the nightly sky, will become operational. Equally important, ``big-data'' algorithms are increasingly being applied and developed to manage the large amount of data provided by these facilities. The discovery space opened by rare or peculiar transients is very large, involving all categories of sources. For low or high redshift objects, a highly efficient UV spectrograph can shed light on a variety of physical ingredients and, in this context, the possible synergy of the CUBES and other ESO facilities at Paranal could open exciting perspectives.

## 2.4 Composition and activity of small bodies of the solar system

Small bodies of the solar system are crucial to understand how our solar system formed and evolved. The gas coma of comets contains a large number of emission features in the near-UV, which are diagnostic of the composition of the ices in its nucleus and the chemistry in the coma. Production rates and relative ratios between different species reveal how much ice is present and inform models of the conditions in the early solar system. CUBES will lead to advances in detection of water from very faint comets, revealing how much ice may be hidden in the main asteroid belt, and in measuring isotopic and molecular composition ratios in a much wider range of comets than currently possible, provide constraints on their formation temperatures. The near-UV is also extremely interesting to study the atmosphere and surface of objects in the outer solar system, including the Galilean and Saturnian satellites, Triton and several Kuiper Belt Objects (KBOs). Measurements performed with CUBES will be essential to complement data obtained by the ESA JUICE mission and the NASA Europa Clipper mission.

## 2.5 From Science to Top Level Requirements

The science cases of interest for the CUBES community have been used to flow-down the Top Level Requirements (TLR) in Phase-A and to effectively contribute in the design trade-offs, via use of software tools developed in the study (ETC, E2E simulator[2,4,7]), both in the previous project phases (Phase A and Phase-B) and at the beginning of the current Phase-C. Some of the Key TLRs, identified for the development of the instrument architecture and design, are:

- Spectral range: CUBES shall provide a spectrum of the target over the entire wavelength range of 305–400 nm in a single exposure (goal: 300 – 420 nm).

- Efficiency: The efficiency of the spectrograph, from slit to detector (included), shall be > 40% for 305–360 nm (goal > 45%, with > 50% at 313 nm), and > 37% (goal 40%) between 360 and 400 nm.

- Resolving power (R): In any part of the spectrum, R shall be > 19000, with an average value > 20000. – R is defined as the full width at half maximum (FWHM) of unresolved spectral lines of a hollow cathode lamp in the spectral slice.

- Signal-to-noise (S/N) ratio: In a 1 hour exposure the spectrograph shall be able to obtain, for an A0-type star of U = 17.5 mag (goal U 18 mag), a SNR = 20 at 313 nm for a 0.007 nm wavelength pixel (at an airmass of 1.16). For different pixel sizes, the SNR shall scale accordingly.

An important development in the Phase A study was the potential provision of a second (lower) resolving power (with R ~ 7k), to enable background-limited observations of faint extragalactic sources where spectral resolution is less critical. This resolution and operational mode is in the present instrument baseline design, as shown in the following sections.

## 3.  CUBES SYSTEM ARCHITECTURE

In order to meet the key TLR mentioned in section 2.5, the required entrance aperture of 1.5"x10" (for the high resolution HR, while 6"x10" for the low resolution LR) on sky and the mass limit of 2500 kg for Cassegrain instruments at the VLT, CUBES is a 2-arms spectrograph equipped with reflective image slicers and high efficiency transmission gratings (see the optical path overview described in section 6 and details in the specific proceeding[6]); the main optical bench is made of CFRP (Carbon Fiber Reinforced Polymer, see section 5) and the science detectors to cover the full wavelength range and optical sampling are 9k, 10μm pixels (see section 6). The calibration sources as well as their electronics are hosted in the calibration cabinet (see section 7), while two other cabinets are dedicated for the main instrument control electronics and detector and cryo-vacuum control electronics; the layout of the three cabinets around the main optomechanical structure of CUBES was selected to stay within the allowed volume under the VLT Cassegrain focus.

Being a Cassegrain Instrument, CUBES will be affected by gravitational flexure according to the telescope orientation during science observations. In addition, also thermal effects are relevant since CUBES is not a temperature stabilized instrument like the ones operating in a (vacuum) vessel. The required instrument stability is a spectral stability of 0.5 pixels (on average over the wave-band) over 24 hours and zenith distances from 0 to 60 degrees. This is driven by the goals of minimizing R degradation during long integrations, relatively high RV stability and maximizing observing efficiency by avoiding attached calibrations. This is difficult to achieve for an instrument which has a quite large beam size and via only purely passive mechanical stiffness, even in static condition at zenith angles larger than 0. Therefore, CUBES design foresees the implementation of active flexure compensation, described in section 9, in order to cope with differential flexures between the slicer output focal plane (the spectrograph "equivalent entrance slit") and the detector plane (the spectrograph focal plane). CUBES will use the recently released ELT control software framework. Communication with the VLT infrastructure (CUBES will operate at VLT in Cerro Paranal) is granted by the VLT / ELT gateway software component.

The main CUBES active functions, spread over the different identified subsystems, are depicted in Fig.1; these are:

- *Calibration sources switch on/off (including CCD Flat-Field) and filters options*: to operate the different calibration operations (i.e. the common calibrations template and frames for proper data reduction)

- *Calibration light injection and pin-holes masks selection*: used for calibration and instrument characterization purposes

- *Acquisition and Guiding path*, fed by the motion of the A&G mirror, which allows for the implementation of a specific Imaging operational mode (in addition to the operations for science object acquisition and secondary guiding, on top of the telescope guiding/tracking)

- *Main path and A&G ADCs:* rotation to correct for atmospheric dispersion during observation

- *Exposure shutter:* controlled by the NGC-II (ESO New Generation Controller – II)

- *Image slicers (HR and LR) selection:* to enable different spectral resolutions and modes

- *Collimators motion for active flexure compensation (AFC):* actuators to decenter the collimator lens (one per each arm)

- *Camera focusing:* to compensate for focus variation due to changing temperature, implemented at the beginning of each scientific observation
- *Cryo-Vacuum:* to implement the Detectors pressure and temperature operative conditions inside the Cryostats

At system level, the identification of instrument linear and rotational motion active functions drove the definition of requirements in term of mass to be moved, motion range and speed, repeatability for motor and stages selection.

In Figure 1, the functions are represented via icons and geometries, following the light path from the telescope Cassegrain interface (top) to the detector end (bottom). Subsystem/product decomposition (described later in this section) is indicated by dashed regions.

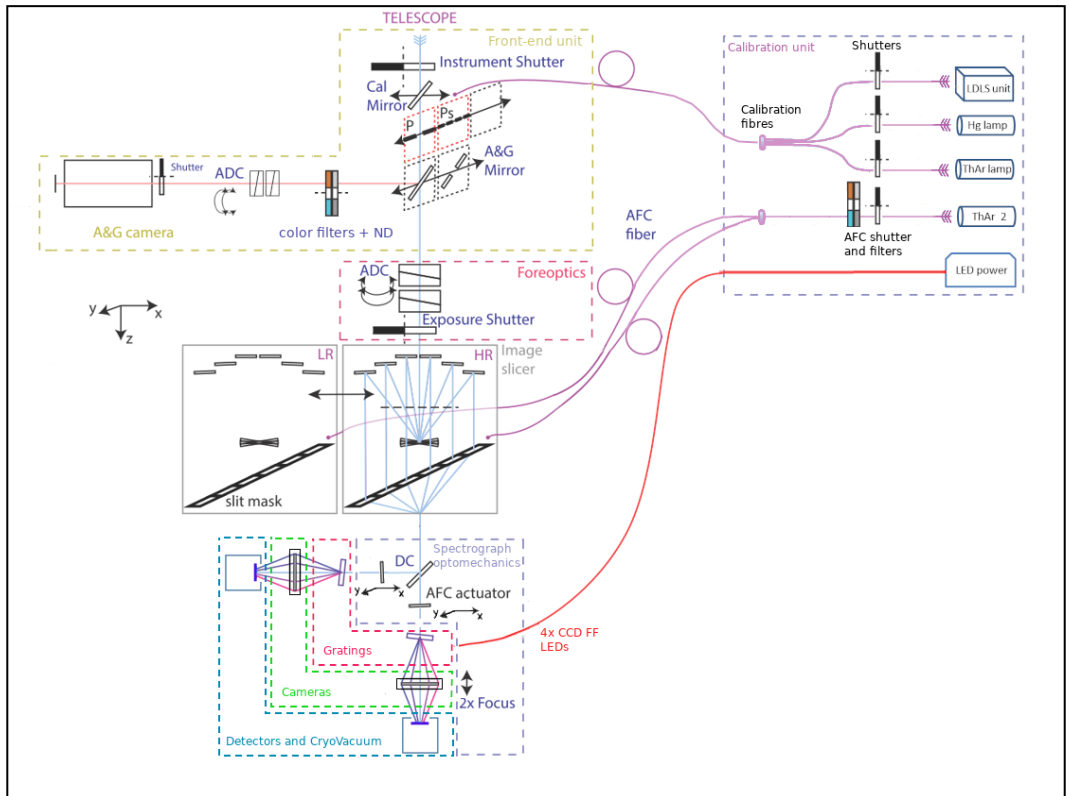

Figure 1. CUBES architecture and functional diagram. Functions are represented via icons and geometries, Subsystem/product decomposition (described later in this section) is indicated by dashed regions.

As with other VLT instruments, CUBES operational model envisages a mixture of Visitor Mode and Service Mode observations, with the transient cases also requiring Rapid-Response-Mode (RRM) or Target-of-Opportunity (ToO) Observations. A stand-alone CUBES will be offered with HR and LR configuration mode. An imaging mode will also be offered with the acquisition camera to provide photometric calibrations of the targets. It is not foreseen to use the Imaging mode alone.

A long process has been devoted to defining the different physical subsystem as first level of the product breakdown structure, according to an approach oriented to the Manufacturing Assembly Integration and Test phase (MAIT) but without forgetting the functional decomposition common practice. Such product oriented approach has the purpose to aid the project development, different MAIT aspects (both technical and managerial like assemblies readiness reviews) and subsystems interfaces definition and control. Table 1 gives a very brief description of the CUBES subsystems.

Table 1. CUBES Subsystems general description

| Subsystem | General Description |
|---|---|
| Front-End Unit | Top bench tower including all items inside. Calibration-Unit injection assembly. A&G camera: from first lens to TCCD, including filter wheel and all opto-mechanics. |
| Calibration unit | Cabinet with internal functions and LCUs, calibration sources. Fibers up to I/F or connectors. |
| Fore-Optics: | From fore-optics collimator lens (first optical item of fore-optics optical path) to exposure shutter (before the slicer) including all optics and opto-mechanics. Including rotation mechanisms and motors for (main path) ADCs. |
| Bench | Main optical Carbon-Fiber bench, including top-side and bottom-side covers. |
| Spectrograph Opto-Mechanics | From slit to collimated beam, opto-mechanical elements. Including actuation mechanism and motors for collimator XY motion and cryostat I/F brackets. |
| Slicer | For both HR and LR: Slicer mirrors, slicer camera mirrors, slit mask, baseplate, cover. AFC projector for HR and LR. Including linear stage and motor for Slicer(s). |
| Camera(s) | Optics and mechanical barrel of both cameras. Including stages and motor of both cameras. |
| Grating(s) | Dispersive elements, including mounts and mechanical frames. |
| Global structure | Support frame assembly, telescope adapter, cable wrap, and possible counterweights. |
| Handling tools | Handling tools, alignment tool (if any) and Transport container(s). |
| Detector & Cryo-Vacuum | Cryostats with Detector devices inside. Cryo-Vacuum components (incl. control electronics and cabinet) and NGCII. Cooling of Detector and Cryo-Vacuum cabinet (i.e. manifold, circuit and control). |
| Main Instrument Control Electronics | Main ICE cabinet, ICP, cables up to devices. NOTE: motors distributed among other Subsystems. |
| Cooling | Pipes and routing from SCP to ICE and CAL Cabinets and to subsystems (A&G TCCD). Except for DET cabinet. |
| Science software | Data Reduction Software, Exposure Time Calculator. |
| Instrument control software | Function Control software, Observation Control software, Templates. |

Model Based System Engineering (MBSE) approach has been implemented to support some of the system engineering tasks, mainly: requirements management, activities modeling, and generation of system structure documents, like the Product Breakdown Structure (PBS) or the Bill of Materials (BoM). Requirements management controls the flow-down process to have a coherent list of requirements. This is achieved using derived properties and tailored numbering. Activities modeling uses traditional MBSE techniques while mimicking the software templates for calibration and observation. To generate structural documents, the system structure is generated in Cameo using the outputs from the subsystems, granting high coherency between the model and the actual design state. The interaction with non-Cameo users relies on Excel files, accessible to all interested parties and usable by Cameo to export and import information. Detailed description of MBSE techniques implementation in CUBES can be found in a parallel proceeding[5].

## 4. CUBES OPTICAL DESIGN

The instrument optical path will consist of (see details here[6]):

- Front-End subsystem which has the task of:
  - o Injecting the light from the Calibration unit into the main path – done with a combination of two COTS off-axis parabola.
  - o Providing acquisition and guiding functionalities. The acquisition and guiding camera is fed by a fold mirror placed after the telescope focus. The A&G optical path (shown in Figure 2) is based on two counterrotating 3-prism ADCs, ensuring correction over the FoV > 90x90 arcsec and 300-700 nm range. RMS spot radius is < 150 mas, ADCs residuals are <200 mas. A filter wheel allows to insert u,g,r SDSS and Neutral Density filters used for photometric calibration in the imaging mode. The A&G detector is 1K x 1K, 13 µm pixels format.

- o Inserting 1) a pin-hole mask: single pin-hole 1-arcsec size used for alignment or 2) a row of 7 pin-holes 0.25 arcsec size for PSF characterization along slit or 3) a clear position that lets the science beam pass-through the fore-optics path.

- Fore-Optics: providing a collimated beam for the ADCs (which correct over the design wave-band 300-405 nm for zenith angle up to 60° as per specifications) and magnifying optics to feed the image slicer. The total aperture is 6x10 arcsec. All refractive optics are made of fused silica or CaF2.

- High Resolution Image slicer[3] re-formats the rectangular 1.5 x 10arcsec FoV generating the spectrograph entrance slit; the "equivalent entrance slit" is composed by six slit-lets, each one of 0.25 x 10 arcsec at ≈F/19.41 to the spectrograph. The LR Image Slicer has a FOV of 6 x 10 arcsec. Both slicers are composed by two sets of mirrors: slicer mirrors and slicer-camera mirrors; these are made in Fused-Silica with aluminum coating (foreseen for the former) and dielectric coating (foreseen for the latter) considered to optimize efficiency.

- The spectrograph optical path includes (see Figure 3):
  - o A dichroic beam splitter to spectrally separate the full bandwidth into 2 arms (reflecting the "Blue" passband - 300-351.6 nm - and transmitting the "Red" passband – 345.6-405 nm).
  - o Then for each arm:
    - ▪ Two folding mirrors
    - ▪ A collimating lens (used also to implement AFC); the nominal collimated beam size 160 mm.
    - ▪ A transmission grating to provide the necessary spectral dispersion; the groove density is 3597 lines/mm and 3115 lines/mm. The required first order diffraction efficiency is >88.5% and >89.5% (blue and red arm goal, at the AOI. S-P polarizations, wavelength and surface average over clear aperture, including reflection losses at the A/R coated side)
    - ▪ A camera to focus the spectrum on the detector.

The optical layout was optimized such that all optical elements of the spectrograph from slit to detector lie in a single plane, so all spectrograph optics can be mounted on a single optical bench (with size 1.3×1.7 m2).

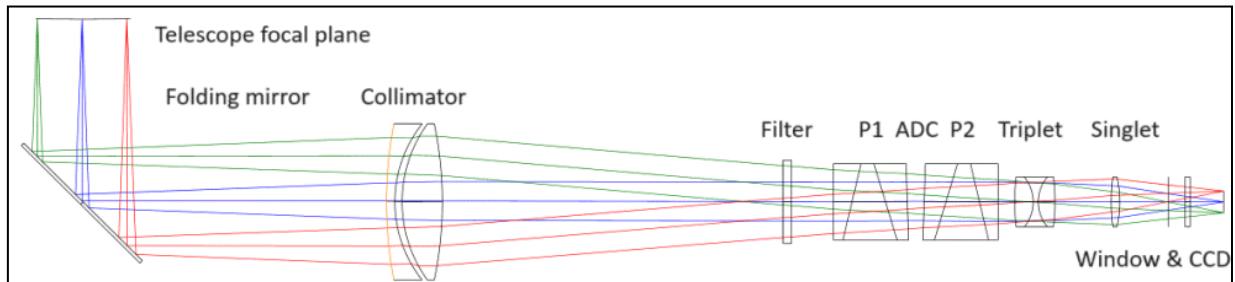

Figure 2. Optical design layout of the CUBES A&G camera, specific components mentioned in the figure.

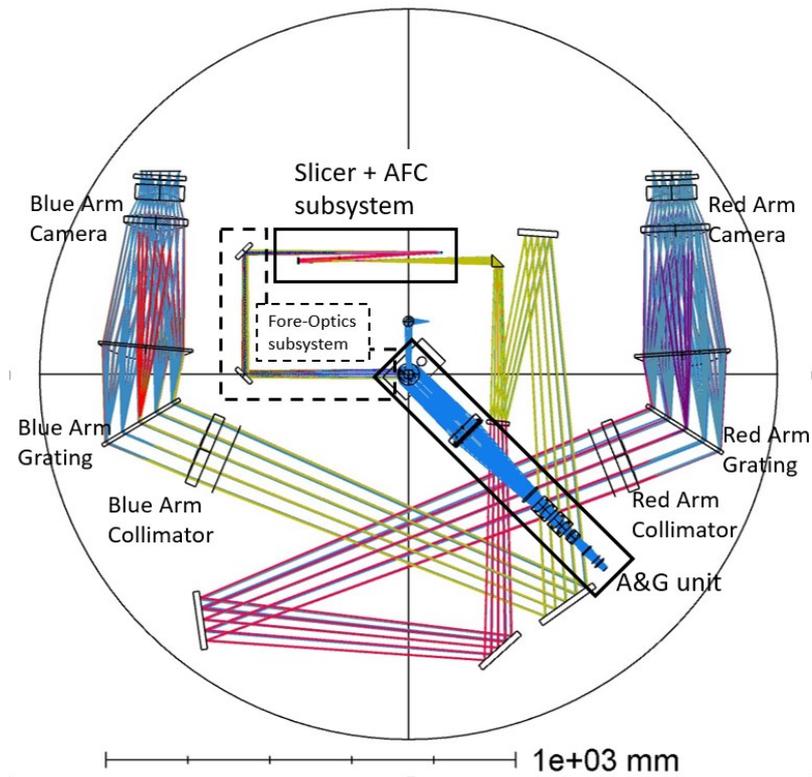

Figure 3. Optical design layout of CUBES full path: A&G camera, Calibration injection unit, Fore-optics, Slicers (including AFC projector unit), 2-arms spectrograph with specific components mentioned in the figure.

The RMS spot diagram of the blue arm, for different wavelengths and central field of each slice is shown in Figure 4. More details can be found in a parallel proceeding[6].

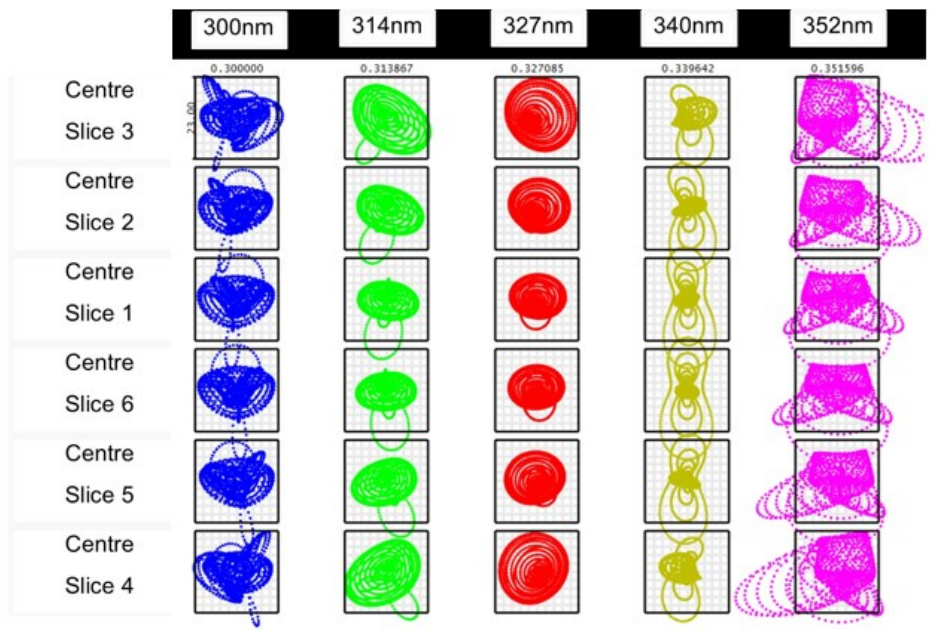

Figure 4. RMS spot diagram of the blue arm, for different wavelengths and central field of each slice. Box size is 23 μm.

# 5. CUBES MECHANICAL DESIGN

As explained in section 0, all optical elements of the spectrograph from slit to detector lie in a single plane, so all spectrograph optics can be mounted on a single optical bench (with size 1.3×1.7 m2). This is arguably the most stable configuration since the dispersion direction of CUBES is parallel to the stiff surface plane of the optical bench.

CUBES requires a fairly large beam diameter of 160 mm. Consequently, the instrument envelope is also rather large compared to other Cassegrain instruments (cf. X-Shooter with a 100 mm beam diameter). Scaling classical instrument designs to the required size of CUBES would result in exceeding the mass limit of 2500 kg for UT (Unit Telescope) Cassegrain instruments. Therefore, a light-weight construction principle has been adopted, making use of modern composite materials, in particular for the main optical bench and the A&G sub-bench.

The major mechanical assemblies are:

- The optical bench assembly (see Figure 5), that provides a stable platform for the spectrograph optics as well as for the fore-optics and front end. It is a 1.3×1.7 m2 carbon fiber reinforced polymer (CFRP) bench. The A&G sub-bench, which is part of the Front-End subsystem, is also made of CFRP to minimize CTE differences between the two benches.

- The global structure (see Figure 6), which provides mechanical support for the three cabinets mounted with custom bars and brackets, is made of steel S355, and connects the main optical bench to the Cassegrain flange via a truss-structured telescope adapter. A single adapter ring at the top interfaces to the Telescope Cassegrain flange. It also includes, as subsystem definition, the cable wrap structure and gives mechanical support to pumps and vacuum equipment.

The opto-mechanical mounts will be made of standard aluminum alloy, all optics are glued into their metal housings using room-temperature-vulcanizing silicon (RTV). Appropriately dimensioned bonding gaps guarantee an athermal behavior of the mounts to minimize thermal stress on the optical elements. As a Cassegrain instrument, CUBES sees varying gravitational loads during operation. The CFRP main Bench and all optomechanical mounts are designed to provide the required stiffness and distribute loads minimizing high stress regions. Finite Element Analysis is used to verify the design approach and estimate the displacements of optical elements at varying telescope positions (see Figure 5).

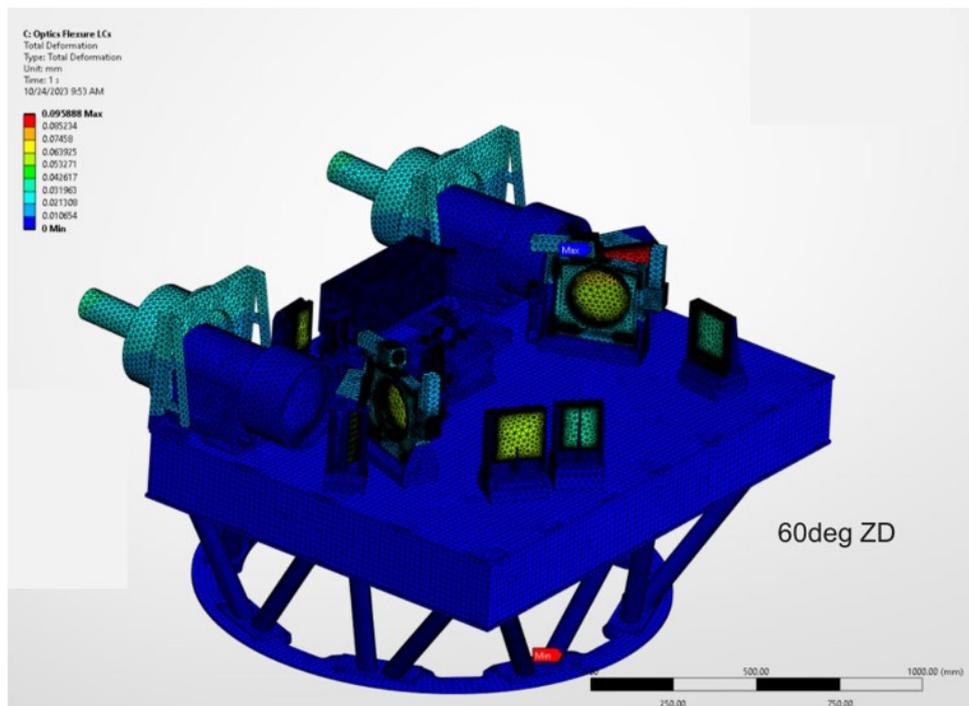

Figure 5. Flexure analysis of CUBES CFRP Bench, optomechanics and telescope adapter.

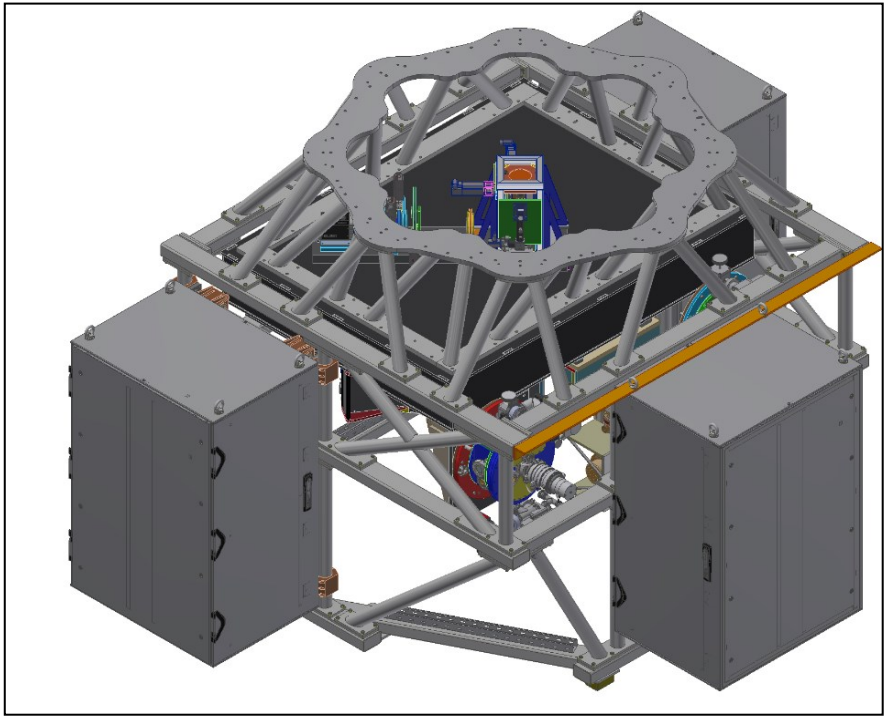

Figure 6. CUBES instrument, including global structure support frame & cabinets. Covers are removed.

# 6. CUBES DETECTOR AND CRYO-VACUUM

The two cryostats provide the operational conditions (T=165 K) to the selected science detector CCD290-99, 9k, 10μm pixels. This is the best option to fulfil different TRS (e.g. Efficiency, Sampling, Resolving power). The selected coating for both CCDs is the ML25 proposed by the manufacturer. Each CCD will be controlled by one single ESO NGC-II module; NGC-II is able to read the two halves of CCD independently to allow for the implementation of the AFC functionality. The cryogenic CCD pre-amp is currently being prototyped to ensure environmental robustness.

The CUBES detector cryostat is based on a heritage design provided by ESO, modified to implement specific features for CUBES (see Figure 7). Copper links conduct heat from the cold mass to the selected cryocooler, the Sunpower GT AVC, which provides substantial cooling power and negligible exported vibration. Vacuum design (i.e. pumps, valves, pressure sensors) is based on ESO guidelines.

The interface flange to the bench bracket provides 6 degrees of freedom adjustment.

The cryostat control system uses Siemens S7 PLC, integrating the sorption pump and main copper link heater. A Lakeshore 336 with DT670 sensors controls the science detectors set-point.

The layout of electronic and control modules inside the detector cabinet (as well as for the Main ICE and Calibration cabinets) is under development to allow for optimized air flow circulation.

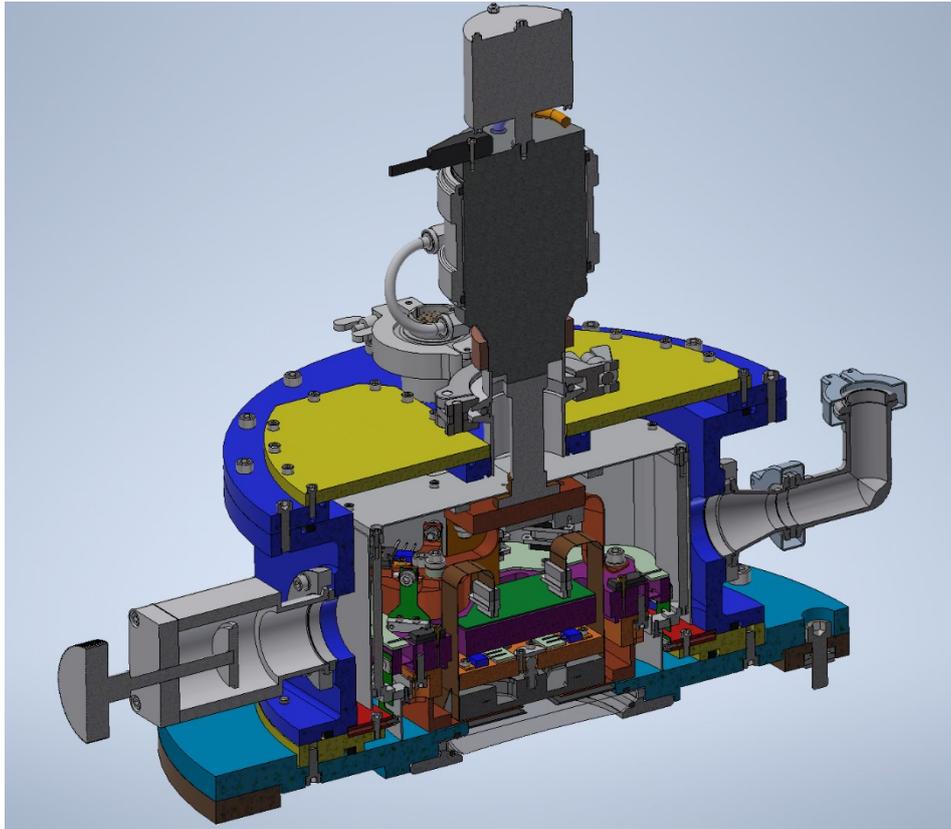

Figure 7. Cryostat mechanical design, with details of I/F flange, inner CCD mount & read-out electronics, cryocooler with AVC.

## 7. CUBES CALIBRATION UNIT

The calibration unit (CU) provides the calibration lights: flat fielding (LDLS and LEDs), wavelength calibration (ThAr lamp), alignment (Hg lamp), flexure correction (ThAr lamp). All lamps and optical feeding lines to the optical fibres stay onto a calibration lamp bench installed in a cabinet drawer (see Figure 8), which will have a custom Divinicel light/dust tight cover. On the other side of this bench, all calibration unit electronics and control modules (based on Beckhoff modules as per ESO standards) are mounted. An additional full cover to encapsulate the whole bench in order to avoid light leakage is under final evaluation.

Calibration Unit cabinet hosts all the lamps, optical feeding lines to the optical fibres and related control electronics devices (see Figure 8). The lamp and electronics bench is accessible from both side doors of the cabinet and can slide if unit extraction is needed for maintenance.

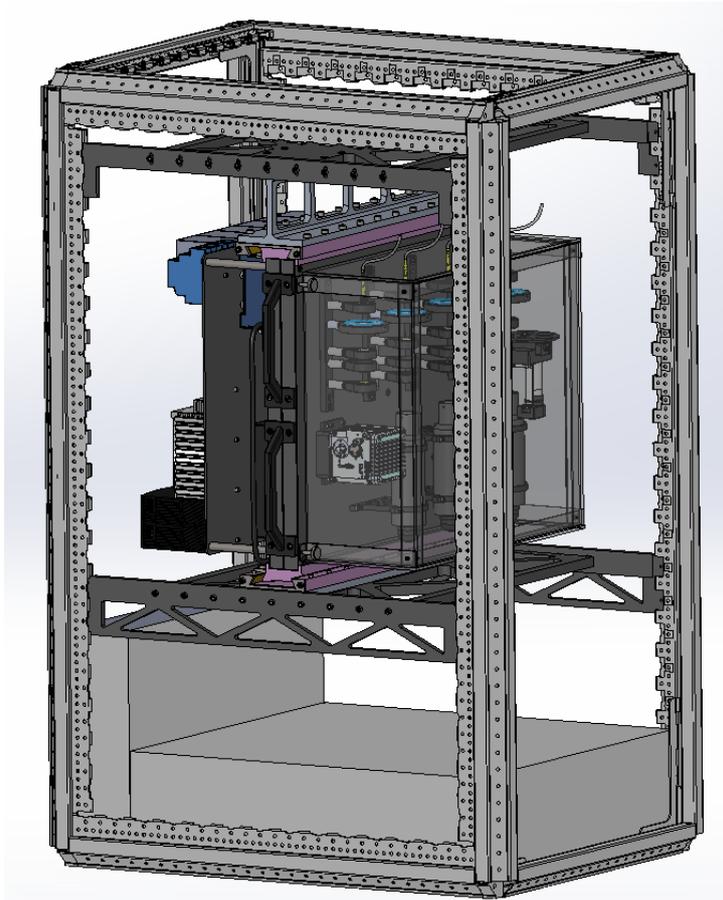

Figure 8. Calibration-Unit Cabinet. Frame with the sliding structure which holds the lamps and electronics bench. Lamps on the right side, electronics on the left side. Cabinet doors not shown in the figure.

## 8. CUBES SOFTWARE

The CUBES instrument benefits from an entire "software ecosystem" built around it, whose individual packages cooperate to support the users from the proposal preparation to the science-grade spectra. Relevant components are:

- OPS (Observation Preparation Software): to collect all required information and prepare the observations;
- ETC (Exposure Time Calculator): used to estimate the exposure time required to achieve a given Signal-to-Noise Ratio for a specific exposure;
- ICS (Instrument Control System): to control all the instrument functions and detectors, as well as to coordinate the observations;
- DRS (Data Reduction Software): to calibrate science exposures and extract 1D spectra;
- E2E (End-to-end Simulator): to simulate raw images without the need for a physical instrument, allowing early characterization of the performances as well as early tuning of the DRS (see details here[7]).

Although operating at VLT, CUBES will be one of the first instruments adopting the recently released ELT control software framework and related standards, which will be used to develop and control all the future ESO/ELT instruments. A specific software component, dubbed VLT / ELT gateway, is foreseen to allow communication between the CUBES software and the VLT environment. Further information on the CUBES software ecosystem are available here[4].

## 9. CUBES ACTIVE FLEXURE COMPENSATION (AFC)

As underlined in section 3, to cope with thermo-gravitational flexures of Cassegrain and not-temperature stabilized instrument, CUBES implements an AFC system/functionality to stabilize the spectral traces onto the CCD with respect to the nominal position to better then 0.5 pixels (on average over the wave-band) over 24 hours; this specification is driven by the goal of minimizing R degradation and ensuring spectral format stability to maximize observing efficiency and avoid attached calibrations.

The light of a dedicated ThAr (Thorium-Argon) lamp from the Calibration Unit is injected via a dedicated unit shown in Figure 10, called AFC projector, at the slicer assembly (one for the HR slicer and one for the LR slicer). It projects the light-beam at the output focal plane of the image slicer with an F/# around 18. A slit of 0.2 mm x 0.75 mm will be located at the output focal plane and mounted directly on the opto-mechanics of the AFC-projector. This ensures the best control over the stray-light, however a demanding control on the slit position must be provided. The optics of the AFC projector will be mounted within a tube fixed to a rectangular mounting plate. For each image slicer, a block shall be permanently connected to the INVAR (tbc) baseplate of the image slicer. The AFC projector will be screwed onto the INVAR baseplate and three pins will be used for alignment (by shimming) and will ensure repeatability. See all details in Figure 10.

From the AFC slit, the light is projected by the spectrograph onto the top half of the CCD, while the science spectrum falls on the bottom half as shown in Figure 9. This is processed with a dedicated software algorithm (see details here[7]) which computes the shift of the AFC spectrum (both in spectral and spatial directions) with respect to an AFC reference spectrum taken during daytime with the telescope at zenith.

The XY corrections are actively performed by moving the Collimators which are installed into flexure mounts (Figure 11).

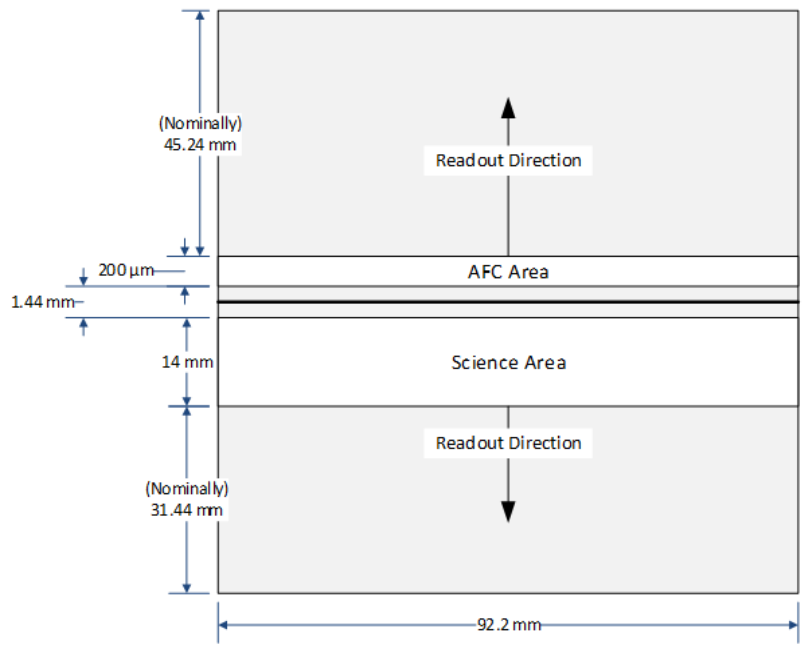

Figure 9. CUBES Used detector regions: the top half of the CCD is used to collect the AFC spectrum, while the science spectrum is projected by the optical path of the instrument onto the bottom half.

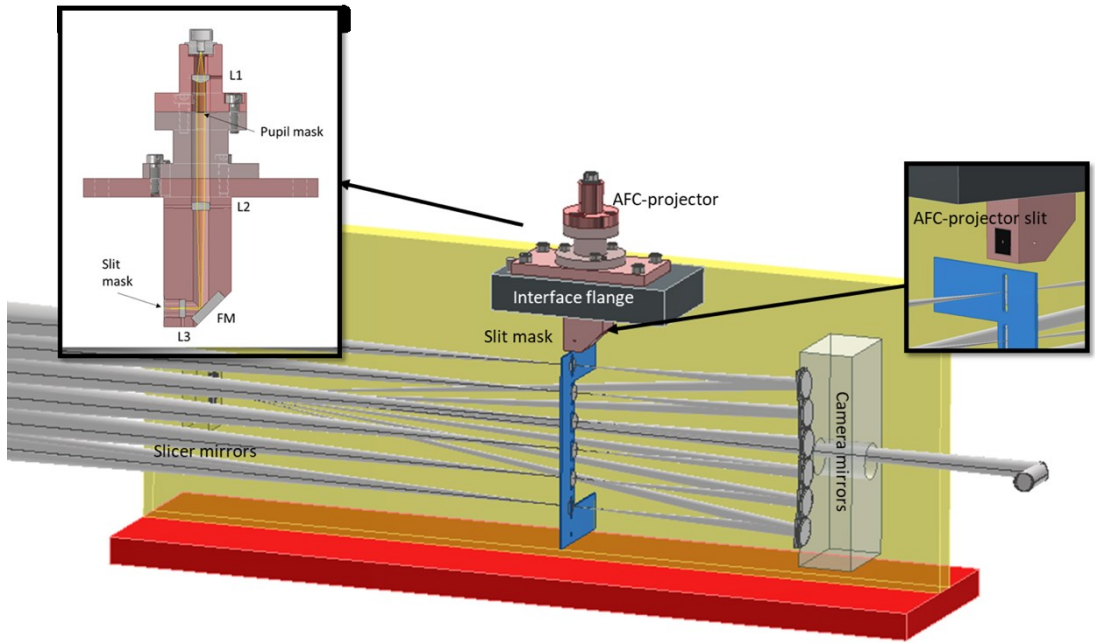

Figure 10. AFC projector Optomechanical layout. The yellow surface represents the image slicer baseplate for the slicer mirror, camera mirrors and slit mask (blue unit). The dark-gray INVAR block is the interface flange for the AFC projector.

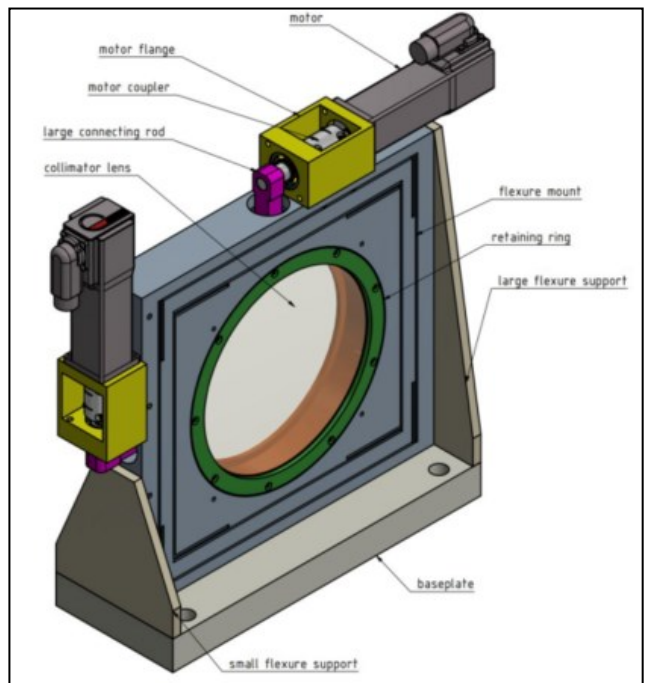

Figure 11. AFC active XY motion implementation on Collimators via motors and flexure mounts.

# 10. CUBES ASSEMBLY INTEGRATION AND TESTING

The Assembly, Integration, and Testing (AIT) in Europe is divided between Italy and Germany as sketched in Figure 12.

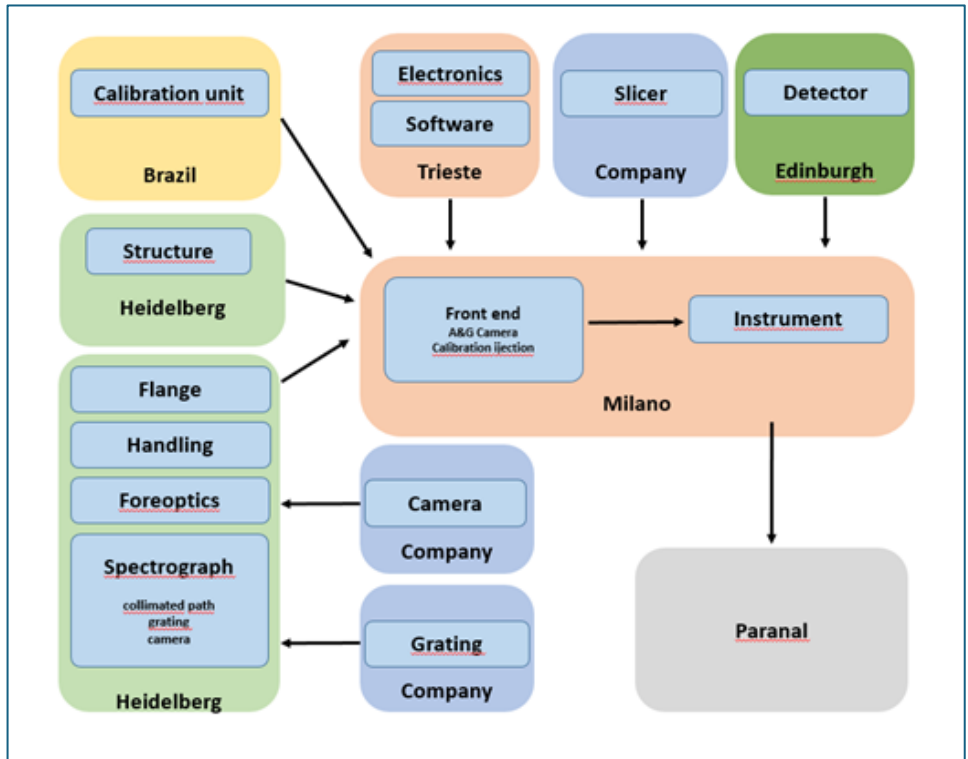

Figure 12. Scheme of the movements of the different subsystem in Europe.

The overall process is divided into two phases:

**First Phase:**

- Location: Landessternwarte (LSW), Zentrum für Astronomie der Universität Heidelberg, Germany.
- Activities: The bench will be accepted and mounted onto a dedicated handling tool. Fore-optics, spectrograph, and camera will be pre-integrated in Heidelberg. Since the slicer is not available during this phase, the fore-optics and spectrograph will be tested separately with technical detectors.

**Second Phase:**
- Location: INAF-Osservatorio Astronomico di Brera, Italy.
- Activities: Assembly of the remaining subsystems will occur here. The mechanical axis of the interface flange will be used as a reference to align the other subsystems (Front-end and Fore-optics). The detector will then be integrated and aligned with the camera assembly. Next, the slicer will be integrated, linking all the subsystems together. Meanwhile, the calibration unit will be connected to the system, and its light will be used to test the entire instrument.

During these integration phases, the activities will be performed with the bench mounted on a dedicated handling tool that allows the system to be flipped upside down, ensuring proper accessibility for handling all components. Afterward, the external structure is assembled, and the bench is fixed onto it along with the cabinets.

**Final Testing:**

- Location: Merate.(Italy)
- Activities: All final system tests will take place on a Cassegrain simulator, where the effects of different orientations can be tested in a configuration that simulates the telescope's mechanical environment.

## 11. PROJECT MANAGEMENT AND PRODUCT & QUALITY ASSURANCE

### 11.1 The CUBES Consortium

The CUBES Consortium is composed by institutes from five countries:

1. **INAF** - *Istituto Nazionale di Astrofisica*, Italy (Consortium leader)

2. **STFC-UKATC** - *UK Astronomy Technology Centre*, (primary UK partner) and **Durham** *University Centre for Advanced Instrumentation* (secondary UK partner), United Kingdom

3. **LSW** - *Landessternwarte, Zentrum für Astronomie der Universtität Heidelberg*, Germany

4. **NCAC** - *Nicolaus Copernicus Astronomical Center of the Polish Academy of Sciences*, Poland

5. **IAG-USP** - *Instituto de Astronomia, Geofísica e Ciências Atmosféricas* (primary Brazil partner) and **LNA** - *Laboratório Nacional de Astrofísica (secondary Brazil partner)*, Brazil

Each country is represented in the managerial structure by one PI (Principal Investigator) or Co-PI. The PI of the leading technical institute, INAF, is the single point of contact between ESO and the Consortium for all contractual matters.

### 11.2 Project Schedule

After an one-year Phase A conceptual design completed in June 2021, and after the endorsement by the ESO Council at the end of 2021, the CUBES Preliminary Design Phase (Phase B) started in February 2022 with the signature of the Construction Agreement between ESO and the leading institute of the CUBES Consortium. CUBES adopts the standard project phasing for ESO instruments which is based on the stage-gate paradigm.

The Preliminary Design Review (end of 2022) has been followed by the Detailed Design Phase (Phase C), with the Final Design Review foreseen in October 2024. During the Phase C, the CUBES Consortium successfully completed the Long Lead Items Reviews of the Detector (July 2023) and Optics (November 2023), that allowed to start the purchase processes of the instrument items that are critical for instrument performance, have long delivery times and are technologically challenging (Science Detectors, Slicers, Gratings and Spectrograph Optics). The current project schedule foresees the Preliminary Acceptance Europe (PAE) in June 2028 and the Preliminary Acceptance Chile (PAC) in September 2029.

### 11.3 Product and Quality Assurance

According to the ESO Common Requirements, the CUBES Spectrograph must reach dependability and safety goals during its life. To achieve this objective, continuous iteration between the Product and Quality Assurance team (PA/QA) and the designers, with a strong contribution of the system engineering, is established.
The PA/QA activities have regarded the 3 typical areas:

- The Product and Quality Assurance plan definition, agreed at the beginning of the phase B and kept updated according to the needs of the project.
- The Reliability, Availability and Maintainability analysis[8], carried out in both the quantitative and qualitative ways (RBD, functional analysis, and FMEA), in order to prove the dependability of the System according to the requirements.
- The Hazard analysis, performed according to the international norms.

The aforementioned analyses have demonstrated that CUBES is a reliable instrument and the hazardous situation during the PAE, PAC and operational phases have been identified, discussed and along with proposed mitigation procedures[9]. The following table summarize the results of both the analysis. Updates will be released at the end of the phase C.

Table 2. Synoptic MTBF analysis table for CUBES

| Sys MTBF Budget = 17280h | Severity | n. RLE | RLE MTBF budget [h] | RLE out of budget |
|---|---|---|---|---|
| CALIBRATION UNIT | 3 | 7 | 3.63E5 | 0 |
| DETECTOR & CRYO-VACUUM | 2 cryog. | 2 | 1.94E4 | 0 |
| | 3 cryog. | 4 | 2.59E4 | 0 |
| | 1 | 2 | 3.46E4 | 0 |
| | 2 | 2 | 5.18E4 | 0 |
| | 3 | 1 | 5.18E4 | 0 |
| ELECTRONICS & COOLING | 1 | 24 | 4.15E5 | 6 |
| | 2 | 7 | 1.81E5 | 3 |
| | 3 | 27 | 1.41E6 | 11 |

# 12. KEY DESIGN PERFORMANCE

## 12.1 Wavelength Coverage

The CUBES design approaching the FDR guarantees to cover a wavelength range from 300 to 405 nm, thus comfortably larger than the target requirement of 305-400 nm band. Specifically, the first arm is 300-351.6 nm, while the second arm 345.6-405 nm; the overlap of 6nm allows for the transition from reflection to transmission of the dichroic.

## 12.2 Resolving Power

The (spectral) resolving power is computed as: $R_\lambda = \frac{\lambda}{LSF_{\lambda,FWHM} \times \Delta\lambda}$. Where the $LSF_{\lambda,FWHM}$ is the FWHM of the Gaussian fit of the spectral line spread function, which in turn is the convolution of:

a) The width on the detector of an image of each slicer mirror. This is defined by the physical aperture of each slice.
b) The spectral intensity distribution of aberrations generated by the optics.
c) Diffusion of the signal in the detector between pixels.
d) Movement of the spectra on the detector due to flexure and temperature changes.
e) Detector flatness.

According to our analysis, the diffusion of the signal in the CCD290-99, 9k detector between pixels is characterized by a Gaussian with a full-width half maximum of 0.9 pixels. Thermal and flexure opto-mechanical offsets are modelled as a Gaussian with a full-width half maximum of 0.5 pixels. The detector has a peak-to-valley flatness specification of 20 microns, to model the effect of this on the spectral resolution a ±10-micron defocus applied to the position of the detector. This essentially gives an absolute worst-case value for the resolving power degradation due to this defocus at all wavelengths. The resulting spectral Resolving Power is shown in Figure 13. The figure shows both the minimum R which occurs in the different slice, and the mean value of the 6 slices. R is above the minimum specification of 19000 for the full wavelength range in both arms; also the average along the wave-band is above the required 20000.

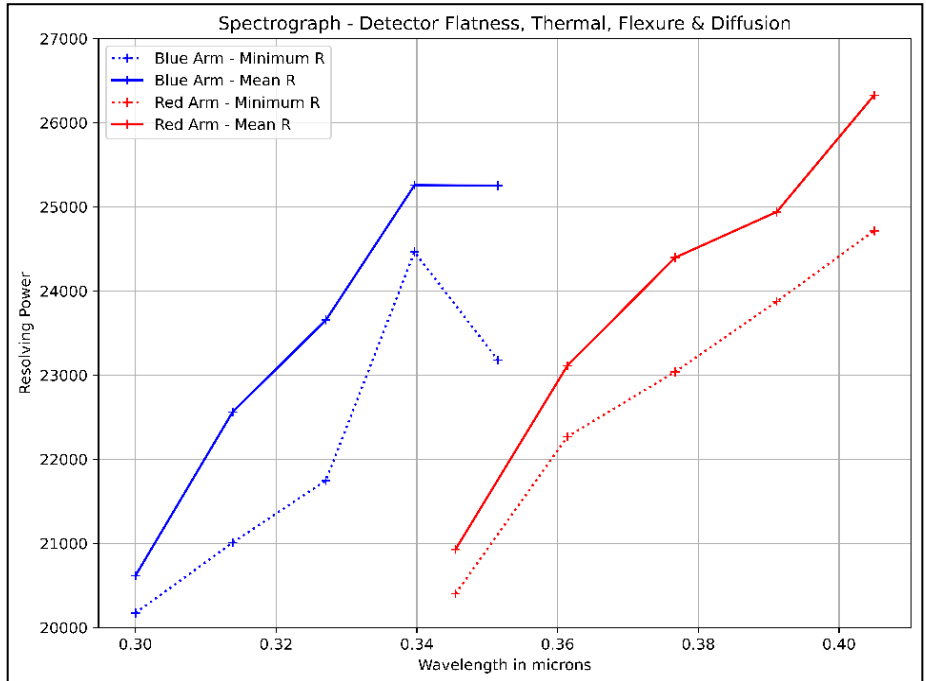

Figure 13. CUBES computed (High Resol.), R > 20000. Minimum per slice, and average on the 6 slices.

## 12.3 Efficiency

Efficiency estimations from previous phases (phase-A and PDR) were made considering two possible cases related to performance of AR coatings, vignetting, cement and gratings. Detailed inquiries to the most promising coating suppliers for AR and HR coatings in the CUBES wave-band has been done in order to have realistic detailed achievable values. Two scenarios have been computed: *goal and worst case*. It must be underlined that in the *goal* case both design values (for AR, HR, Dichroic – and CCD coating) and "typical/goal" values (cements, vignetting) the manufacturer commit to be able to meet are considered; while in the *worst* case the average minimum specification the manufacturers commit to be able to meet are considered for each type of optical element.

Figure 14 shows the goal and worst case scenarios for both arms. The goal/typical case is well above minimum specifications for both arms. The worst-case scenario is also above specification except for the blue edge (below 310 nm) and the red edge (above 397 nm); in both edges the worst case overall performance is below specifications of just <1%.

It is again underlined that these two scenarios shall be regarded as the envelope within which the final CUBES instrument efficiency performance is expected to fall, thus being above specification with a good probability figure.

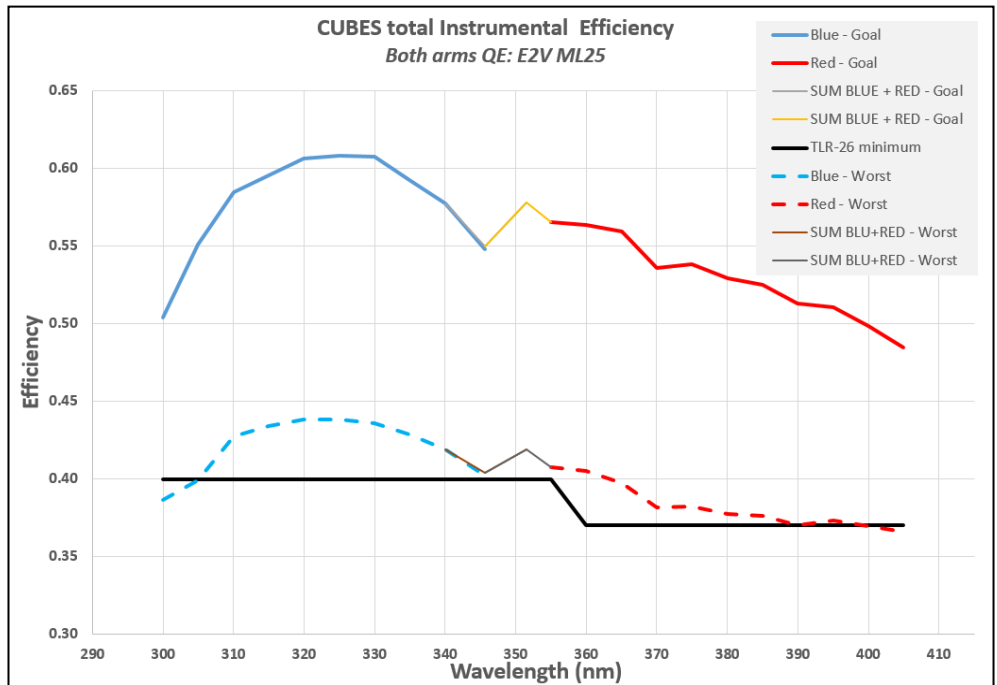

Figure 14. CUBES estimated Instrumental efficiency. Design/Goal and Worst case scenarios, based on manufacturers coatings transmission design and minimum guaranteed data.

### 12.4 Signal to Noise Ratio

The requirement asks for: *"In a one-hour exposure time the spectrograph shall be able to obtain for an A0 star magnitude U>=17.5 (goal U>=18) a SNR =20 for a wavelength pixel of 0.007nm at 313 nm (at an airmass of 1.16). For different pixel sizes and airmass value, the S/N ratio shall scale accordingly."*

A reference spectrum of U=17.5 in AB mag system is assumed to be observed at standard sky conditions at seeing 0.87 arcsec ($\lambda$=500nm); VLT telescope collecting area and throughput are properly taken into account, while Slicer-slit losses as well as number of pixels in spatial direction are computed according to the seeing and plate scale.

For what concerns Dark Current (DC), RON and optical efficiency values two cases have been simulated, for the calculation of the achievable SNR:

- The "goal-case", where:

  o the detector DC = 0.5 e-/pix/hr, at the CUBES operative temperature of 165K,

  o CCD Readout Noise (RON), taken from read noise graphs from the detector datasheet, in Single ended readout @100 kHz is 2.5 e- rms,

  o the instrument efficiency set as a typical-design goal scenario (see plot in Figure 14);

- The "conservative-specification case" ("worst-case"), where:

  o the detector DC (3 e-/pix/hr, at temperature of 173K)

  o CCD Readout Noise (RON), taken from read noise graphs from the detector datasheet, in Differential readout @100 kHz is 3.1 e- rms,

  o All the efficiency values represent the most conservative expectations, i.e. performing (<u>all at the same time</u>) at the average minimum specification manufacturers commit to be able to reach (see dashed line in Figure 14).

The SNR is computed in a classical way, i.e. no optimal extraction is simulated, per *a wavelength pixel of 0.007nm at 313 nm* and in binning 1x1; the SNR in goal-case scenario is about 28, while in the worst-case scenario it is 20, therefore CUBES performance it is expected to fall into this range and being compliant with the requirement.

# 13. CONCLUSIONS

This proceeding presented the status of the Cassegrain U-Band Efficient Spectrograph (CUBES) for the ESO VLT, towards the final design review (FDR). The key science cases were briefly summarized. The instrument operational modes and hardware architecture, to fulfill the science goal and top level requirement, has been described. The different subsystems, identified following a MAIT oriented approach, were presented mentioning system modeling and management through MBSE.

The optical and mechanical design of the full instrument, including detector cryostats, cryo-vaccum subsystem and the calibration unit, was presented highlighting relevant design choices: spectrograph optical path in a single plane, CFRP bench and athermal mechanical mounts, low-vibration cryocooler and accessibility of calibration lamps for maintenance.

The AFC functionality design, to ensure the stability of the spectral traces onto the CCD with respect to the nominal position to better then 0.5 pixels (on average over the wave-band) over 24 hours, was detailed.

The AIT plan is summarized, mentioning the main phases and assembly/integration sites in Germany and Italy. A brief overview of the software ecosystem, project management and quality assurance method was given.

Analysis of the key performance shows that it will deliver outstanding ($> 37\%$) throughput across its bandpass (300-405 nm in the present design, exceeding the 305-400 nm TLR), at a mean R $\sim$ 24000 (HR mode) and R $\sim$ 7000 (LR mode). With contributions from institutes in five countries, the CUBES design is well placed to deliver the most efficient ground-based spectrograph at near-UV wavelengths, opening unique discovery space for the VLT for years to come.

# 14. ACKNOWLEDGEMENTS

R.S. acknowledges support by the Polish National Science Centre through project 2018/31/B/ST9/01469.